\documentclass[onecolumn,iop,apj]{emulateapj}   

\pdfoutput=1

\usepackage{graphicx}        % For eps figures, newer & more powerful
\usepackage{amssymb}         % useful mathematical symbols
\usepackage[usenames,dvipsnames]{color}          % For color text: \color command

\DeclareFontFamily{OT1}{pzc}{}
\DeclareFontShape{OT1}{pzc}{m}{it}%
             {<-> s * [1.1500] pzcmi7t}{}
\DeclareMathAlphabet{\mathscr}{OT1}{pzc}%
                                 {m}{it}

\usepackage{amsmath}
\newcommand{\half}{{\textstyle\frac{1}{2}}}

\newcommand{\pderiv}[2]{\frac{\partial#1}{\partial#2}}
\newcommand{\pderivd}[2]{\frac{\partial^2#1}{\partial#2^2}}

\newcommand{\e}{\hat{\mathbf{e}}}

\newcommand{\B}{{\mathbf{B}}}

\newcommand{\vdot}{{\boldsymbol{\cdot}}}

\newcommand{\grad}{\mbox{\boldmath$\nabla$}}
\newcommand{\bxi}{\mbox{\boldmath$\xi$}}

\newcommand{\thth}{\hspace{1.5pt}}

\newcommand\Div{\grad\vdot\thth}

\newcommand{\ri}{{i}}

\renewcommand{\leq}{\leqslant}  \renewcommand{\le}{\leqslant}

\allowdisplaybreaks[1]

\begin{document}

%%%%%%%%%%%%%%%%%%%%%%%%%%%%%%%%%%%%%%%%%%%%%%%%%%%%%%%%%%%%%

\title{Benchmarking Fast-to-Alfv\'en Mode Conversion in a Cold MHD Plasma. II.\\ How to get Alfv\'en waves through the Solar Transition Region}

\author{Shelley C.~Hansen and Paul S.~Cally}

\affil{Monash Centre for Astrophysics and School of
Mathematical Sciences,\\ Monash University, Clayton, Victoria 3800, Australia}
  \email{shelley.hansen@monash.edu}  \email{paul.cally@monash.edu}

\shortauthors{S.C. Hansen and P.S. Cally}

\shorttitle{Fast-to-Alfv\'en Mode Conversion II}

\begin{abstract}
Alfv\'en waves may be difficult to excite at the photosphere due to low ionization fraction and suffer near-total reflection at the transition region (TR). Yet they are ubiquitous in the corona and heliosphere. To overcome these difficulties, we show that they may instead be generated high in the chromosphere by conversion from reflecting fast magneto\-hydro\-dynamic waves, and that Alfv\'enic transition region reflection is greatly reduced if the fast reflection point is within a few scale heights of the TR. The influence of mode conversion on the phase of the reflected fast wave is also explored. This phase can potentially be misinterpreted as a travel speed perturbation, with implications for the practical seismic probing of active regions.
\end{abstract}


%%%%%%%%%%%%%%%%%%%%%%%%%%%%%%%%%%%%%%%%%%%%%%%%%%%%%

\section{Motivation}

In recent years, Alfv\'en waves have been shown to be ubiquitous in the solar corona \citep{TomMcIKei07aa,McIde-Car11aa} though these may more properly be classed as magneto\-hydro\-dynamic (MHD) kink waves where coronal magnetic structuring supports such tube waves \citep{EdwRob83aa,VanNakVer08aa}. In either case, they are essentially transverse and incompressive, and may be referred to as ``Alfv\'enic''.  Alfv\'en waves have also been detected \emph{in situ} in the solar wind \citep{BelDav71aa}, of which they are postulated to be crucial drivers \citep{Hol06aa}. Furthermore, Alfv\'en waves are inferred in the upper chromosphere based on Hinode Solar Optical Telescope (SOT) observed transverse oscillations of spicules \citep{De-McICar07aa}. Although long-period Alfv\'en waves (hours) dominate observations in the solar wind \citep{BelDav71aa,Hol06aa}, coronal observations reveal oscillations largely in the 100--500 s range, with a discernible peak at 3--4 mHz apparently associated with p-modes. In this paper we shall be concerned only with these shorter period waves, as our main purpose is to explore the extent to which coronal Alfv\'en waves may be extensions of the Sun's p-mode wave field. The source of long-period waves must be sought elsewhere.

Traditionally, it has been thought that direct generation at the solar photosphere by granular buffeting is responsible for the Alfv\'en waves observed further out in the solar atmosphere and beyond \citep[e.g.][]{cravan05aa}. However, this does not take account of the {inefficiency of photospheric MHD Alfv\'en generation \citep{Par91aa,Col92aa}, nor of low ionization fraction effects (\citealt{VraPoePan08aa}, though see \citealt*{TsaSteKop11aa} for a contrary view)}.\footnote{Of course, Alfv\'en waves may be produced in isolated structures in the low atmosphere.  For example, \citet{JesMatErd09aa} identify torsional Alfv\'en waves in a bright-point group of kilogauss strength, though the height of excitation is not clear from this study, and ionization fraction may be enhanced compared with quiet Sun.}

Another impediment to Alfv\'en waves reaching the corona is the strong reflection they suffer at the chromosphere-corona transition region (TR; \citealt{UchSak75aa}). This effect can even produce a trapped-wave resonance structure \citep{SchCalBel84aa}, although as pointed out by \cite{cravan05aa} it is somewhat detuned by non-isothermal stratification. The resonances are also completely absent in the case of an open lower boundary as adopted here. \citeauthor{cravan05aa} quote typical Alfv\'en reflection coefficients of around 95\%, and we concur in general though perhaps we would put that figure even higher at around 98\% in the cases discussed here. Nevertheless, \citeauthor{cravan05aa} conclude that this may still be enough to supply chromospheric and coronal energy losses and to power the solar wind, contrary to the view of \citet{RosLowHol86aa}. 

Photospheric Doppler and Zeeman observations at two heights \citep{Ulr96aa} suggest that indeed there may be sufficient Alfv\'en flux in the 5-minute band to supply coronal losses in the quiet Sun. Indirect arguments suggest that even the order-of-magnitude greater energy requirements of active regions may be consistent with these observations, though this conclusion relies on extrapolation. The velocity data are strongly indicative of outgoing Alfv\'en waves, though the noisier magnetic data may suggest the addition of a reflected (downgoing) component, which might be expected given the high reflectivity of the transition region.

Irrespective of this, the TR is undoubtedly a considerable hurdle for Alfv\'en waves, and it remains to be seen whether the observed coronal Alfv\'en amplitudes are consistent with waves originating in the photosphere alone in both quiet and active regions. Is there some way that the Alfv\'en transmission coefficient can be increased?\footnote{The situation in coronal loops may be different, with resonances allowing greatly enhanced transmission \citep{Hol91aa}. We focus on open field only though.}

We postulate and explore a novel possibility: that the Alfv\'en waves have not propagated as transverse waves (Alfv\'en or kink) from the photosphere, but that instead they have been generated by local mode conversion from reflecting fast waves high in the chromosphere \citep{Mel77aa,MelSim77aa,CalGoo08aa,CalHan11aa,KhoCal12aa}, where low ionization fraction is not an issue for the 3--6 mHz frequency range of most interest here. It will be shown that this greatly increases Alfv\'en penetration of the TR, thereby opening up a potentially more fruitful source of coronal Alfv\'en waves. 

The progenitor fast waves bear the signature of this process not only in their partial loss of energy to the Alfv\'en waves but also in their phase as they return to the interior. Local helioseismology uses phase to infer wave travel times \citep[e.g.,][]{DuvJefHar93aa,BraLin00ab,GizBir05aa}, so it is of interest to gauge this anomalous effect. This is discussed in Section \ref{phase}.

%%%%%%%%%%%
\section{Fast-to-Alfv\'en Conversion: Reprise of Paper I}\label{reprise}

The solar interior is populated by p-modes, global modes of oscillation excited by convection near the surface and trapped in a frequency-dependent cavity between an acoustic cutoff at the top and the Lamb depth at the bottom. At frequencies above about 5 mHz the acoustic cutoff barrier is breached and the waves can escape into the solar atmosphere. Strong inclined magnetic fields, as found in sunspots and other smaller magnetic field concentrations, can reduce the effective cutoff to allow even lower frequency acoustic waves to escape \citep{JefMcIArm06aa}.

Now, in magnetic regions the Alfv\'en speed $a$ increases rapidly with height due to density stratification. So at some point, inevitably, the Alfv\'en speed surpasses the sound speed $c$.
As the waves pass through the Alfv\'en-acoustic equipartition layer where $a$ and $c$ coincide, mode conversion can occur \citep{SchCal06aa,Cal07aa}, splitting the waves into fast (i.e., predominantly magnetic) and slow parts. As the slow wave propagates higher it becomes progressively more field-aligned and more acoustic in nature. Conventionally, the resultant slow waves are 
said to be caused by \emph{transmission} and the fast modes by \emph{conversion}. The amount of transmission/conversion here is dependent on the attack angle $\alpha$, the angle between the wavevector and magnetic field direction.  A small attack angle allows greater transmission to the slow wave and correspondingly a larger attack angle produces greater conversion to the fast wave. In an exact isothermal model conversion to the fast wave can be near total at large $\alpha$ \citep{HanCal09aa}.  

Here we assume that this process has taken place, and follow the fast wave as it progresses to ever higher Alfv\'en speeds $a\gg c$. In this regime we adopt the simplifications afforded by the cold plasma approximation and set $c=0$, or equivalently $\beta=0$ where the plasma-$\beta$ represents the ratio of gas to magnetic pressures.

In Paper I of this sequence \citep{CalHan11aa} we investigated fast-to-Alfv\'en mode conversion in a simple $\beta=0$ plasma with uniform inclined magnetic field $B_0(\cos\theta,0,\sin\theta)$ and exponentially decreasing density $\rho\propto e^{-x/h}$ in the direction $x$ of inhomogeneity, where $h$ is the scale height.  This may be thought of as the vertical direction in a plane stratified model of the solar atmosphere in which density decreases exponentially with height. A fast magneto\-hydro\-dynamic wave is injected from $x\to-\infty$, propagates to the right (positive $x$ direction) and reflects (classically) at $\omega^2=a^2(k_y^2+k_z^2)$, where $a(x)\propto e^{x/2h}$ is the Alfv\'en speed. An $\exp[\ri(k_yy+k_zz-\omega t)]$ time and transverse space dependence is assumed. The orientation of the wavevector in $y$-$z$ space is arbitrary.

If $k_y\ne0$, the fast wave partially converts to Alfv\'en waves, typically around and beyond the fast wave reflection point. Introducing the dimensionless transverse wavenumber $\kappa=\sqrt{\kappa_y^2+\kappa_z^2}$, where $\kappa_y=k_yh=\kappa\sin\phi$ and $\kappa_z=k_zh=\kappa\cos\phi$, it was found that very significant fast-to-Alfv\'en conversion occurs in various regions of $\kappa$-$\theta$-$\phi$ parameter space. Specifically, it was found that:
\begin{enumerate}
\item The thickness of the conversion region depends sensitively on $\kappa$. For $\kappa\sim1$ it is several scale heights thick, for $\kappa\sim0.2$ it extends some 20 scale heights beyond the reflection point, but for $\kappa\sim5$ it is less than one scale height wide. For most solar atmospheric waves of interest in the chromosphere, we might expect $\kappa$ to be considerably less than 1, indicating that the Alfv\'en conversion region can effectively fill the chromosphere.
\item For $\phi\lesssim90^\circ$, conversion is to outgoing (upward) Alfv\'en waves, but for $\phi\gtrsim90^\circ$ conversion predominantly produces backward (downward) propagating Alfv\'en waves. This was to be expected based on a consideration of whether the fast wave aligns with the magnetic field before or after its reflection, consistent with the finding of \citet{Mel77aa} that fast-to-Alfv\'en coupling is strongest at small attack angle.
\end{enumerate}
These predictions were amply confirmed by numerical simulations in a more realistic model sunspot by \citet{KhoCal12aa}.

This simple one-layer model crudely represents the solar chromosphere, and takes no account of the transition region (TR) or corona above. Nevertheless, it allowed us to better understand the local conversion mechanism that goes on there without complications arising elsewhere. In this paper though, we extend the model to consist of two exponential layers, representing the chromosphere $a^2\propto e^{x/h}$ and the corona $a^2\propto e^{x/h_{\rm cor}}$, with $h_{\rm cor}\gg h$ and a discontinuous jump between them representing the TR. This will allow us to explore the consequences of Alfv\'en reflection off the TR and truncation of the extended conversion region by the TR.

%%%%%%
\section{Model, Equations, and Numerical Approach}\label{model}

%\textbf{The overwhelming complexity of plasma physics has led most practitioners to progress by making approximations to focus on their specific phenomenon of interest. In the upper solar atmosphere, the Alfv\'en speed $a$ can exceed the sound speed $c$ by a large margin, particularly above active regions. The available MHD wave types under the circumstance $a\gg c$ are the Alfv\'en wave, with dispersion relation $\omega^2=a^2\kpar^2$, the fast wave $\omega^2\approx a^2 |\k|^2$, and the (field-guided acoustic) slow wave $\omega^2\approx c^2\kpar^2$. In this paper our interest lies in the interaction between the fast and Alfv\'en waves, and so it makes sense to `freeze out' the slow wave by formally taking $c\to0$, or equivalently $\beta\to0$, where the plasma beta $\beta=p_{\rm gas}/p_{\rm mag}$ is the ratio of the gas to magnetic pressures. We are therefore placed well above the Alfv\'en}

In sunspot umbrae the Alfv\'en-acoustic equipartition level $a=c$ is typically situated a few hundred kilometres below the photosphere and at about the photosphere in the penumbra \citep{MatSolLag04aa,BorIch11aa}. Around active regions and in more diffuse magnetic environments equipartition is normally identified with the magnetic canopy of the low chromosphere \citep{BogCarHan03aa,FinJefCac04aa}.\footnote{Usually such discussions are couched in terms of the plasma beta equals unity layer, where $\beta=p_{\rm gas}/p_{\rm mag}$ is the ratio of the gas to magnetic pressure. However, $\beta=2c^2/\gamma a^2$, where $\gamma$ is the ratio of specific heats, so the distinction between $\beta=1$ and $a=c$ is almost immaterial.} As explained in Section \ref{reprise}, seismic waves from the solar interior split at $a\approx c$ into slow (acoustic) waves and fast (magnetic) waves that may propagate into the atmosphere above. We are not concerned with the slow wave here, but the fast wave's fate is very interesting as it can suffer a further mode conversion to the Alfv\'en wave (Paper I). 

Seismic waves below about 3--4 mHz (depending on field strength) may not reach the $a=c$ equipartition level before reflecting. A precondition of the study here is that they have passed through $a=c$, and therefore partially mode converted to magnetically dominated fast waves in $a\gg c$.

With this in mind, we now move well above the $a\lesssim c$ region and adopt a uniform-magnetic-field two-isothermal-layer Alfv\'en speed profile
\begin{equation}
a^2 =
\begin{cases}
A^2 e^{(x-x_T)/h} & x<x_T \\
A^2\,(h_{\rm cor}/h) e^{(x-x_T)/h_{\rm cor}} & x>x_T,
\end{cases}  \label{alfprofile}
\end{equation}
crudely representing the {low-$\beta$} chromosphere ($x<x_T$) and the corona ($x>x_T$) separated by a transition region discontinuity. Here $x$ is the vertical coordinate, increasing with height in the solar atmosphere, $h=c^2/\gamma g$ and $h_{\rm cor}=c_{\rm cor}^2/\gamma g$ are the density scale heights in the two regions, $c$ and $c_{\rm cor}$ are the respective sound speeds, $\gamma$ is the adiabatic index, and $g$ is the gravitational acceleration. The factor $A$ is the Alfv\'en speed at the top of the chromosphere. The exponential increase in $a(x)$ results solely from an exponentially decreasing density $\rho(x)$. The density jump across the TR is of course equal to the inverse of the temperature jump, which explains the $h_{\rm cor}/h$ factor in the second line of Equation (\ref{alfprofile}).
Typically, $h_{\rm cor}\gg h$. 

The crude uniform field model is more applicable to large-scale strong magnetic field regions such as sunspots than to network or other structures where discrete flux tubes at the photosphere quickly expand with height in the atmosphere until they abut and thereafter become more uniform. Especially in this latter case, the exponential Alfv\'en speed increase supposed here will be partially ameliorated by the geometric field spread. Nevertheless, the model represents a useful testbed on which to discern general principles.

To focus on the fast-to-Alfv\'en conversion process in the upper solar atmosphere it is convenient to ignore acoustic waves altogether by neglecting the sound speed $c$ compared to the Alfv\'en speed $a$ in the wave equations, the so-called cold plasma or zero-$\beta$ approximation. The underlying equilibrium stratification is unaffected. This device is used commonly in the study of MHD waves in the upper solar atmosphere \citep{NakVer05aa}. There are innumerable papers in the literature that have studied MHD waves in the cold plasma regime, representing the Sun's outer atmosphere, and have adopted nonuniform density distributions to create Alfv\'en speed inhomogeneities, e.g., \citet{Dav87aa},
\citet{RudRob02aa}, \citet{PasWriDe-10aa}, and \citet{Hol90aa} amongst many others. This last reference is closest to our study, and that of \citet{CalAnd10aa}, in that it discusses a fast MHD wave injected into a region of increasing Alfv\'en speed in which it reflects but also transfers energy to an Alfv\'en resonance. The density profile there is linear though instead of our exponential.  The close correspondence of the results of Paper I, which uses the $\beta=0$ approximation, and the simulations of \citet{KhoCal12aa}, which do not, supports the utility and validity of the cold plasma model in this instance.

We also assume we are not in the low-frequency regime of atmospheric gravity waves \citep{StrFleJef08aa}. Gravity-wave-to-Alfv\'en conversion for very low frequency waves ($\sim 1$ mHz, comparable to the Brunt-V\"ais\"al\"a frequency $N$) has been extensively treated elsewhere \citep{NewCal10aa,NewCal11aa}, and is found to be viable only in very inclined magnetic fields. We therefore assume  $\omega^2\gg N^2$ and that buoyancy plays an insignificant role.  %Hence we may neglect the direct effects of gravity provided we retain the density inhomogeneity that causes stratification in the Alfv\'en speed $a=B/\sqrt{\mu\rho}$. We also neglect transverse variation characteristic of loops.

With all this in place then, we adopt the linearized equation governing the oscillations of a cold plasma (c.f., Equation (1) of Paper I)
\begin{equation}
\left(\partial_\parallel^2+\frac{\omega^2}{a^2}\right)\bxi  =-\grad_{\!\text p}\chi\,,                                \label{basiceqn}
\end{equation}
where  $\bxi(x)=\xi_x\e_x+\xi_y\e_y+\xi_z\e_z= \xi_\perp \e_\perp+\xi_y\e_y$ is the plasma displacement and $\chi=\Div\bxi$ is the dilatation. The subscripts `$\scriptstyle\parallel$' and `$\scriptstyle\perp$' denote the parallel and perpendicular (in the $x$-$z$ plane) directions to the uniform magnetic field $\B_0=B_0(\cos\theta,0,\sin\theta)$ respectively.  The `p' refers to the full perpendicular component to $\B_0$, i.e., $\grad_{\rm p}= \e_\perp\,\partial_\perp+\e_y\,\partial_y$.  Equation (\ref{basiceqn}) neatly expresses the role of the fast wave (represented by $\chi$) as a source of Alfv\'en waves (the term in brackets on the left hand side is the pure Alfv\'en operator).

Frobenius and WKB solutions to this equation with exponential Alfv\'en speed were developed in Paper I for $x\to\infty$ and $x\to-\infty$ respectively, and they are again used here. Indeed, the Frobenius solution is utilized throughout the corona. The imposed boundary conditions are that there are no incoming Alfv\'en waves from $x=\pm\infty$, and that the exponentially decaying fast wave solution is selected at $x=+\infty$ with the exponentially increasing solution deprecated. 
Numerical integration from the WKB region to a suitable matching point in the chromosphere and from the TR also to this matching point completes the solution. Continuity of $\bxi$ and $\partial\bxi/\partial x$ is applied across $x_T$.

%\subsubsection{On the Hankel Solutions and the Transparency of the Exponential Atmosphere}

It is worth reiterating the results from the Appendices of \citet{CalGoo08aa} and Paper I that the point $x\to\infty$ is a regular singular point of the wave equations for the coupled fast and Alfv\'en waves. In the case $k_y=0$ where the two wave types decouple, exact solutions in closed form are available. For the Alfv\'en wave these are in terms of the Hankel functions $H_0^{(1)}$ and $H_0^{(2)}$, representing respectively downgoing and upgoing waves (see  Equation (\ref{hankel}) and Appendix \ref{uni} for full details). In the general case addressed here ($k_y\ne0$), no such closed form solutions are available, but full Frobenius series solutions were constructed in Appendix A.2 of Paper I. These allow us to match our numerical solutions to the pure outgoing Alfv\'en wave at $x\to+\infty$. We note that \citet{cravan05aa} find that reflection of short period Alfv\'en waves is weak in the corona proper, supporting our selection of a radiation boundary condition.

Assuming the incoming fast wave from $x=-\infty$ carries unit wave-energy flux $\mathscr{F}^+=1$, we calculate the outgoing fast flux $\mathscr{F}^-$ at $x=-\infty$, the outgoing Alfv\'en flux $\mathscr{A}^+$ at $x=+\infty$, and the outgoing Alfv\'en flux $\mathscr{A}^-$ at $x=-\infty$. Of course, $\mathscr{A}^+ + \mathscr{A}^- + \mathscr{F}^- = \mathscr{F}^+=1$, which provides a non-trivial test on our numerics.

Determining the relative proportions of the three outgoing modes is our prime goal. In the single layer model of Paper I, any Alfv\'en wave launched in the positive $x$ direction from the conversion region continued outward to yield $\mathscr{A}^+$, whereas an initially backward-propagating Alfv\'en wave excited by the reflected fast wave gave $\mathscr{A}^-$. Here though the picture is complicated by Alfv\'en reflection from the TR, leading us to expect a greatly diminished $\mathscr{A}^+$. Whether $\mathscr{A}^-$ increases or decreases is not clear \emph{a priori}: TR reflected $\mathscr{A}^+$ should enhance $\mathscr{A}^-$, but truncation of the well-spread chromospheric conversion region will diminish it. Numerical solution is required to assess which of these two competing processes prevails.

The full parameter set describing our system is dimensionless wavenumber $\kappa$, magnetic field inclination $\theta$, wavevector polarization $\phi$, transition region position $x_T$, and coronal scale height $h_{\rm cor}$. The chromospheric scale height $h$ is arbitrarily scaled to unity. The frequency $\omega$ and Alfv\'en speed scaling $A$ can be absorbed into the position of fast wave reflection that occurs at $\omega h/a=\kappa$. The zero of $x$ is fixed so that $\omega^2 h^2/a^2=\exp[-x/h]$ in the chromosphere, so fast wave reflection occurs at $x_R=-h\ln\kappa^2$ (provided this is less than $x_T$). We investigate how the various energy fluxes depend on each of the free parameters.

%%%%%%%%%%%%%%%%%%%%%%

\section{Results}

%%%
\subsection{Transmission of the Pure Alfv\'en Wave}
Before embarking on a study of fast-to-Alfv\'en mode conversion, it is useful to first calculate the transmission coefficient $0\le\mathscr{T}\le1$ of a pure Alfv\'en wave incident on the transition region from the chromosphere. Here $\mathscr{T}$ represents the fraction of wave energy flux in the incident Alfv\'en wave that passes into and through the corona. The remainder is reflected at the TR. The pure wave, with $\xi_y$ polarization only ($\phi=0$), is completely decoupled from and therefore not excited by the fast wave, but it does give us a good measure of how strongly the TR reflects Alfv\'en waves. The exact solutions of the Alfv\'en wave equation $\partial^2\xi_y/\partial t^2=a^2\,\partial^2\xi_y/\partial s^2$ (where $s$ is distance along a field line) in an atmosphere with exponential Alfv\'en speed are available in terms of Hankel functions:
\begin{equation}
\xi_y = \Xi_{1,2}\,e^{-i \,k_z (x-x_T)\tan\theta } H_0^{(1,2)}\left( \frac{2\omega h}{a} \sec\theta    \right)\,,
\label{hankel}
\end{equation}
with $H_0^{(2)}$ representing a wave propagating upward and $H_0^{(1)}$ corresponding to downward travel. $\Xi_{1,2}$ are arbitrary amplitudes with dimensions of length. As explained in Appendix \ref{uni}, the Hankel functions represent unidirectional waves. This is consistent with the total Poynting flux in the $x$-direction:
\begin{equation}
F_x = \frac{\omega B_0^2\cos\theta}{\pi\,h\,\mu}\left(\left|\Xi_2\right|^2-\left|\Xi_1\right|^2\right)\,.
\label{Poynting}
\end{equation}

\begin{figure}
\begin{center}
\includegraphics[width=0.48\hsize]{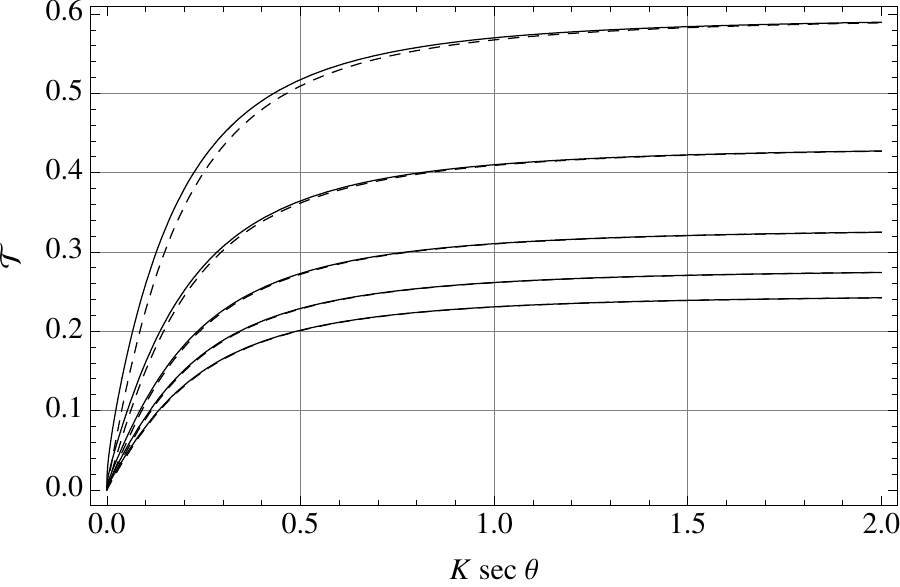}
\hfill
\includegraphics[width=0.48\hsize]{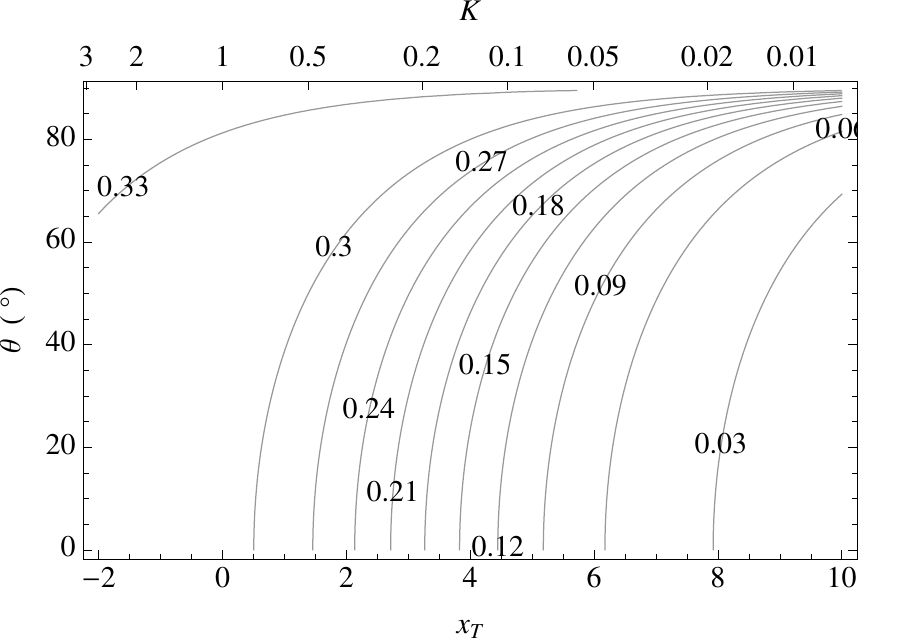}
\caption{Left: Pure Alfv\'en transmission coefficient $\mathscr{T}$ (see Equation (\ref{T})) against $K\sec\theta$ for $h_{\rm cor}=20$, 50, 100, 150, and 200 (full curves, top to bottom). In the limit $h_{\rm cor}\to1$, the TR vanishes and transmission becomes total. The dashed curves represent the transmission coefficient for a uniform corona, as given by Equation (\ref{TU}). Right: Contour plot of transmission coefficient $\mathscr{T}$ against transition region position $x_T$ and magnetic field inclination $\theta$ for the case $h_{\rm cor}=100$. The contour heights are as labelled.}
\label{fig:TpureAlf}
\end{center}
\end{figure}

Matching to an $H_0^{(2)}$ solution (i.e., outgoing only) with the coronal scale height and letting $h=1$ in the chromosphere, the transmission coefficient is
\begin{equation}
\mathscr{T}=
1-\left|\frac{H_0^{(2)}\left(2 K\,h_{\text{cor}}^{1/2}  \sec \theta \right)
   H_1^{(2)}(2 K  \sec \theta )-h_{\text{cor}}^{-1/2}{H_0^{(2)}(2 K  \sec \theta )
   H_1^{(2)}\left(2 K\,h_{\text{cor}}^{1/2}  \sec \theta 
   \right)}}
     {{h_{\text{cor}}^{-1/2} H_0^{(1)}(2 K 
   \sec \theta ) H_1^{(2)}\left(2 K\,h_{\text{cor}}^{1/2}  \sec \theta 
   \right)}-H_1^{(1)}(2 K  \sec
   \theta ) H_0^{(2)}\left(2 K\,h_{\text{cor}}^{1/2}  \sec \theta 
   \right)}\right|^2
   \label{T}
\end{equation}
where $K=\omega h/a(x_T^-)=\exp[-x_T/2h]$ is the dimensionless Alfv\'enic field-aligned wavenumber at the top of the chromosphere. Note that, for fixed $h_{\rm cor}$, $\mathscr{T}$ depends only on $K\sec\theta$. $\mathscr{T}$ increases monotonically from zero at $K\sec\theta=0$ towards
\begin{equation}
\mathscr{T}_{\rm max}=\frac{4\sqrt{h_{\rm cor}}}{(\sqrt{h_{\rm cor}}+1)^2}
\label{Tmax}
\end{equation}
 as $K\sec\theta\to\infty$ (see Fig.~\ref{fig:TpureAlf}). For $h_{\rm cor}=100$, $\mathscr{T}_{\rm max}=0.3306$. Only for $h_{\rm cor}=1$ (no transition region) does $\mathscr{T}_{\rm max}$ reach 1. 
 
 For comparison, the pure Alfv\'en transmission coefficient for the case of a uniform rather than exponential corona is
\begin{equation}
\mathscr{T}=
1-\left|\frac{H_0^{(2)}(2 K \sec \theta )+i\, h_{\rm cor}^{1/2} H_1^{(2)}(2 K
   \sec \theta )}{H_0^{(1)}(2 K \sec \theta )+i\, h_{\rm cor}^{1/2}
   H_1^{(1)}(2 K \sec \theta )}\right|^2\,,
   \label{TU}
\end{equation}
where now $h_{\rm cor}$ represents only the jump in $a^2$ across the TR and not the coronal scale height (which is infinite). It results directly from (\ref{T}) in the limit $2Kh_{\rm cor}^{1/2}\sec\theta\gg1$ where the relevant Hankel functions may be replaced by their exponential asymptotic forms. The two formulae are compared in Figure \ref{fig:TpureAlf}, indicating that the precise choice of coronal structure has little practical significance for the total transmission. Equation (\ref{TU}) reduces to Equation (44) of \citet{LeeHolFla82aa} (for reflection coefficient $\mathscr{R}=1-\mathscr{T}$) in the case $h_{\rm cor}=1$ where a uniform layer is \emph{continuously} appended above an exponential atmosphere.

Interestingly, even with the $h_{\rm cor}=100$ characteristic of the real solar transition region, substantial Alfv\'en transmittance ($>10\%$) is possible if $K\sec\theta\gtrsim0.1$. High frequency, low TR Alfv\'en speed, and large field inclination are all favourable for Alfv\'en wave penetration of the transition region.

However, this is for an incident pure Alfv\'en wave. As shown in Paper I, Alfv\'en waves may also be generated over an extended region by conversion from reflecting fast waves, in fact from the evanescent tail beyond the reflection point. In this interaction region, they are not pure Alfv\'en waves, only becoming so asymptotically as $x\to\infty$. So how do these hybrid waves fare on encountering the TR? And what if the fast waves reach the TR before they can reflect, or early in their evanescent tail? These issues are addressed next.

\begin{figure*}
\begin{center}
\includegraphics[width=\hsize]{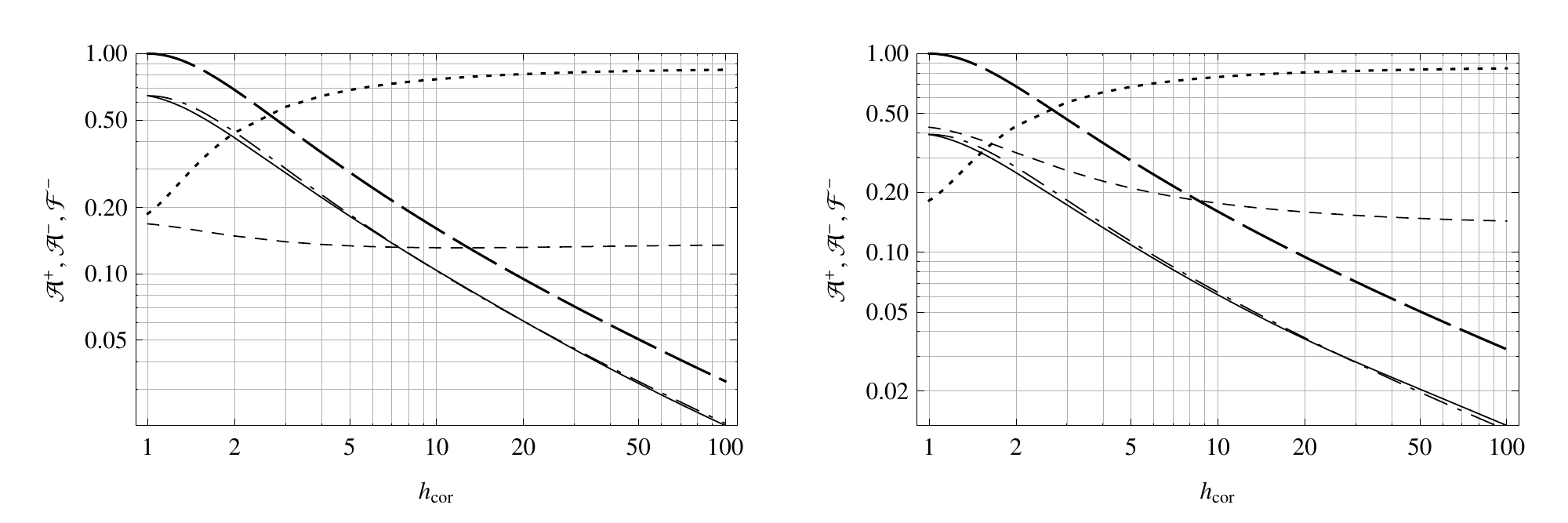}
\caption{The forward Alfv\'en conversion coefficient $\mathscr{A}^+$ (full curve), reverse Alfv\'en conversion coefficient $\mathscr{A}^-$ (dashed curve) and reverse fast conversion coefficient $\mathscr{F}^-$ (dotted curve), as a function of $h_{\rm cor}$ for $\kappa=0.2$ (i.e., $x_R=-\ln0.2^2=3.219$), $\theta=30^\circ$, $x_T=8$, $\phi=70^\circ$ (left panel) and $\phi=110^\circ$ (right panel).  Both variables are plotted on logarithmic scales. The heavy long-dashed curve represents $\mathscr{T}$ as given by Eq.~(\ref{T}). The chained lines represent $\mathscr{A}_0^+\mathscr{T}$, where $\mathscr{A}_0^+(\kappa,\theta,\phi)$ is the forward Alfv\'en conversion coefficient without transition region (see Paper I, or equivalently as calculated here with $h_{\rm cor}=1$).
}
\label{fig:hcor}
\end{center}
\end{figure*}

%%%
\subsection{Alfv\'en Wave Fluxes from Fast Wave Conversion}
The first task is to see how the TR ``jump'' $h_{\rm cor}$ affects the output fluxes (recall that $a^2$ increases across the TR by the factor $h_{\rm cor}/h=h_{\rm cor}$). 
Figure \ref{fig:hcor} shows how the Alfv\'en conversion coefficient $\mathscr{A}^+$ (conversion to the upward propagating Alfv\'en wave), reverse coefficient $\mathscr{A}^-$ (conversion to the downward Alfv\'en wave), reverse fast wave coefficient $\mathscr{F}^-$, pure Alfv\'en transmission coefficient (with TR) $\mathscr{T}$ and $\mathscr{A}_0^+\mathscr{T}$ vary with the coronal scale height $1\le h_{\rm cor}\le100$.   Here we define $\mathscr{A}_0^+$ as the forward Alfv\'en conversion coefficient without the TR.  We fix $\kappa=0.2$ (i.e., $x_R=3.219$), $x_T=8$, and magnetic field inclination $\theta=30^\circ$ with wavevector polarizations $\phi = 70^\circ$ and $110^\circ$.  At $h_{\rm cor}=100$, $K\sec\theta=0.021$ and $\mathscr{T}=0.033$, so a pure Alfv\'en wave is 97\% reflected.

\begin{figure}[htb]
\begin{center}
\includegraphics[width=.45\hsize]{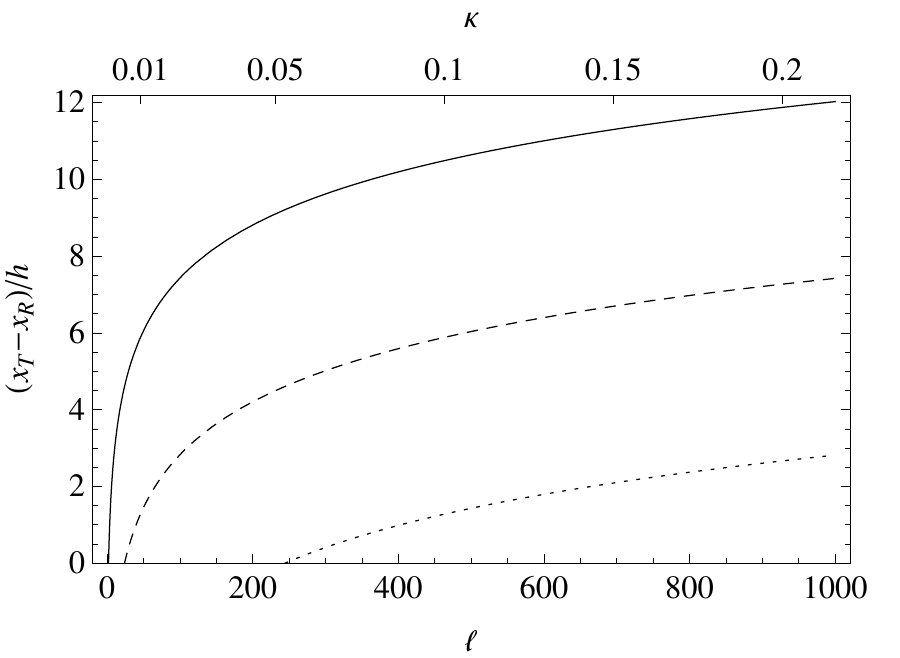}
\caption{Distance $\Delta=(x_T-x_R)/h$ in units of chromospheric scale height between the fast wave reflection height $x_R$ and the transition region height $x_T$ as a function of spherical harmonic degree $\ell$ (or dimensionless wavenumber $\kappa$ on the top axis) and magnetic field strength $B$: Full curve 1 kG; dashed curve 100 G; dotted curve 10 G. The wave frequency is 5 mHz, the chromospheric scale height is 150 km, and the density at the top of the chromosphere is $10^{-10}$ $\rm kg\,m^{-3}$. Where $\Delta$ dips below zero, the fast wave no longer reflects before reaching the TR.}
\label{fig:Delta}
\end{center}
\end{figure}

The distance between the reflection point and the TR, measured in units of the chromospheric scale height $h$, is given by
\begin{equation}
\Delta=\frac{x_T-x_R}{h} = \ln\frac{a(x_T^-)^2\kappa^2}{\omega^2h^2}\,,  \label{Delta}
\end{equation}
where $a(x_T^-)$ is the Alfv\'en speed at the base of the TR. This quantity will be seen to be crucial in determining the Alfv\'en flux penetrating the TR. To place this dimensionless length in a solar context, let us suppose the chromospheric scale height is 150 km, that the density at the base of the TR is  $10^{-10}$ $\rm kg\,m^{-3}$, and that we are dealing with 5 mHz waves. Then, $\Delta$ depends on field strength $B$ and dimensionless horizontal wavenumber $\kappa=\sqrt{\ell(\ell+1)}\,h/R_\odot$. It is plotted in Figure \ref{fig:Delta}, showing that the fast wave nearly reaches the TR at smaller field strength and spherical degree. The latter is due to the wave being more nearly vertical as $\ell$ gets smaller, meaning that it can reach higher in the chromosphere.

The case $h_{\rm cor}=1$ corresponds to the model of Paper I with no transition region. Increasing $h_{\rm cor}$ unsurprisingly produces a sharp decrease, to around 1-2\%, in the amount of $\mathscr{A}^+$ propagating through the corona. Interestingly,  $\mathscr{A}^-$ also decreases, though only modestly. The beneficiary of these decreases is of course $\mathscr{F}^-$, which increases to over 80\% at $h_{\rm cor}=100$, characteristic of the solar TR, meaning most of the injected fast wave flux returns whence it came as a reflected fast wave. Although at $\phi=110^\circ$ the backward Alfv\'en flux $\mathscr{A}^-$ is quite significant in the absence of a TR ($h_{\rm cor}=1$), once $h_{\rm cor}$ reaches 100 there is little difference between the two polarizations, consistent with near-symmetry of forward and backward Alfv\'en production at $\phi=70^\circ$ and $110^\circ$ respectively and the forward Alfv\'en flux having been almost totally reflected.  

Also striking in Figure \ref{fig:hcor} is how well $\mathscr{A}^+$ is mimicked by $\mathscr{A}_0^+\mathscr{T}$. This appears to suggest that the simple picture of Alfv\'en generation according to coefficient $\mathscr{A}_0^+$ followed by TR transmission $\mathscr{T}$ describes the situation remarkably well. This was unanticipated, as the conversion region was not expected to be confined to these few scale heights below the TR.

\begin{figure*}[tb]
\begin{center}
\includegraphics[width=\hsize]{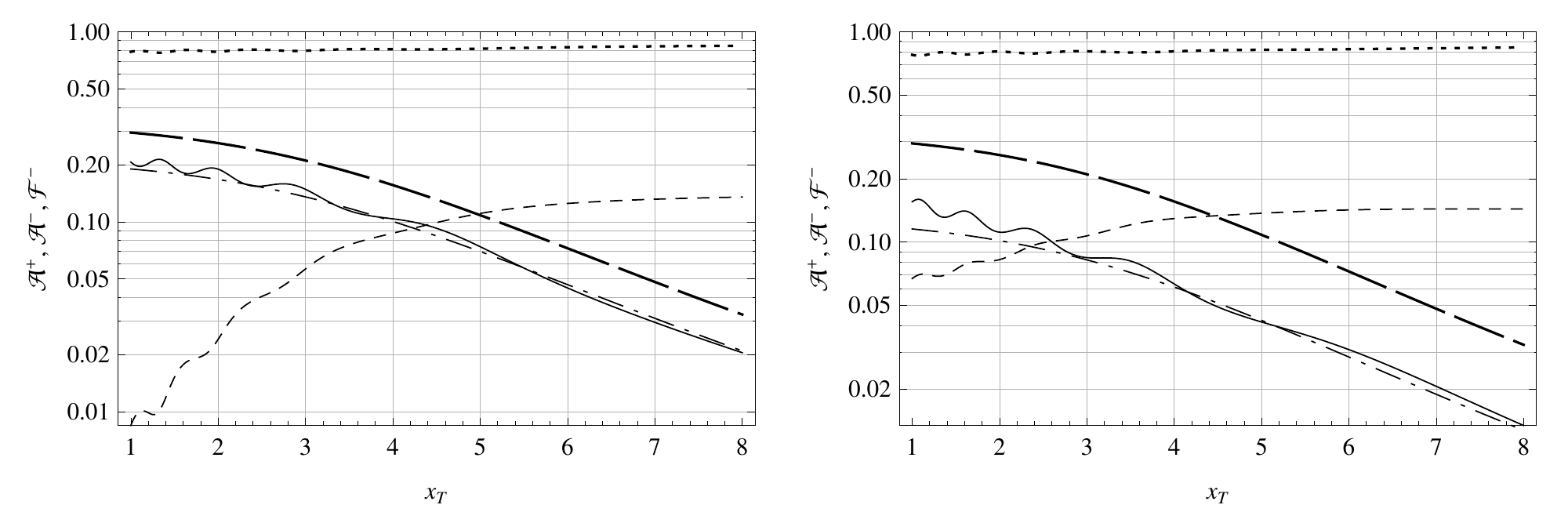}
\caption{The forward Alfv\'en conversion coefficient $\mathscr{A}^+$ (full curve), reverse Alfv\'en conversion coefficient $\mathscr{A}^-$ (dashed curve) and reverse fast conversion coefficient $\mathscr{F}^-$ (dotted curve) as a function of $x_T$ for $\kappa=0.2$, $\theta = 30^\circ$, $h_{\rm cor}=100$, $\phi=70^\circ$ (left panel) and $\phi=110^\circ$ (right panel). The fast wave reflection point is $x_R=-\ln0.2^2=3.219$ if this is less than $x_T$. Otherwise, the fast wave will propagate throughout the chromosphere and immediately be evanescent on entering the corona (provided $\omega^2 h^2/a(x_T^+)^2<\kappa^2$). The heavy long-dashed curve represents $\mathscr{T}$ as given by Eq.~(\ref{T}). The chained lines represent $\mathscr{A}_0^+\mathscr{T}$, where $\mathscr{A}_0^+(\kappa,\theta,\phi)$ is the forward Alfv\'en conversion coefficient without transition region.
}
%The respective maxima in $\mathscr{A}^+$ are: 0.951 for $\theta=10^\circ$ ($\kappa=0.056$, $\phi=88.0^\circ$); 0.814 for $\theta=20^\circ$ ($\kappa=0.107$, $\phi=82.9^\circ$); 
%0.647 for $\theta=30^\circ$ ($\kappa=0.170$, $\phi=73.9^\circ$); 
%0.505 for $\theta=40^\circ$ ($\kappa=1.29$, $\phi=41.0^\circ$).
%}
\label{fig:xT}
\end{center}
\end{figure*}

Next, we fix the magnitude of the transition jump but adjust its distance from the fast reflection point at $x_R$. Figure \ref{fig:xT} demonstrates the dependence of flux on the location of the TR for fixed $\kappa=0.2$, $\theta = 30^\circ$ and $h_{\rm cor}=100$ again for wavevector polarizations of $\phi=70^\circ$ and $\phi=110^\circ$.  As the TR is raised further above the fast wave reflection point, there is a decreasing amount of $\mathscr{A}^+$ with an increase in both downward Alfv\'en and reflected fast wave coefficients. The decrease in $\mathscr{A}^+$ is somewhat surprising as we might have expected that a higher $x_T$ allows a larger
fast-to-Alfv\'en conversion region and therefore potentially more outward Alfv\'en flux, at least for $\phi=70^\circ$ (recall that the conversion region is some 20 scale heights wide at $\kappa=0.2$). This is apparently because the transition region simply becomes more reflective to Alfv\'en waves as the distance between the fast reflection height and the TR increases (see Fig.~\ref{fig:TpureAlf}). Very substantial forward Alfv\'en fluxes $\mathscr{A}^+$ of around 20\% are seen at small $x_T\lesssim x_R=3.219$. It appears that if the fast wave tail penetrates into the low corona, it continues to generate Alfv\'en waves there.

One again we see that $\mathscr{A}_0^+\mathscr{T}$ closely matches the numerical calculation of  $\mathscr{A}^+$.  Of  course as the transition region falls below the fast wave reflection point there are further complications and the two formulae are less well aligned. It may be that the Alfv\'en conversion is largely or partially in the corona in these cases, thereby invalidating the simple interpretation of $\mathscr{A}_0^+\mathscr{T}$ in terms of two sequential processes.

\begin{figure*}[tbhp]
\begin{center}
\includegraphics[width=\hsize]{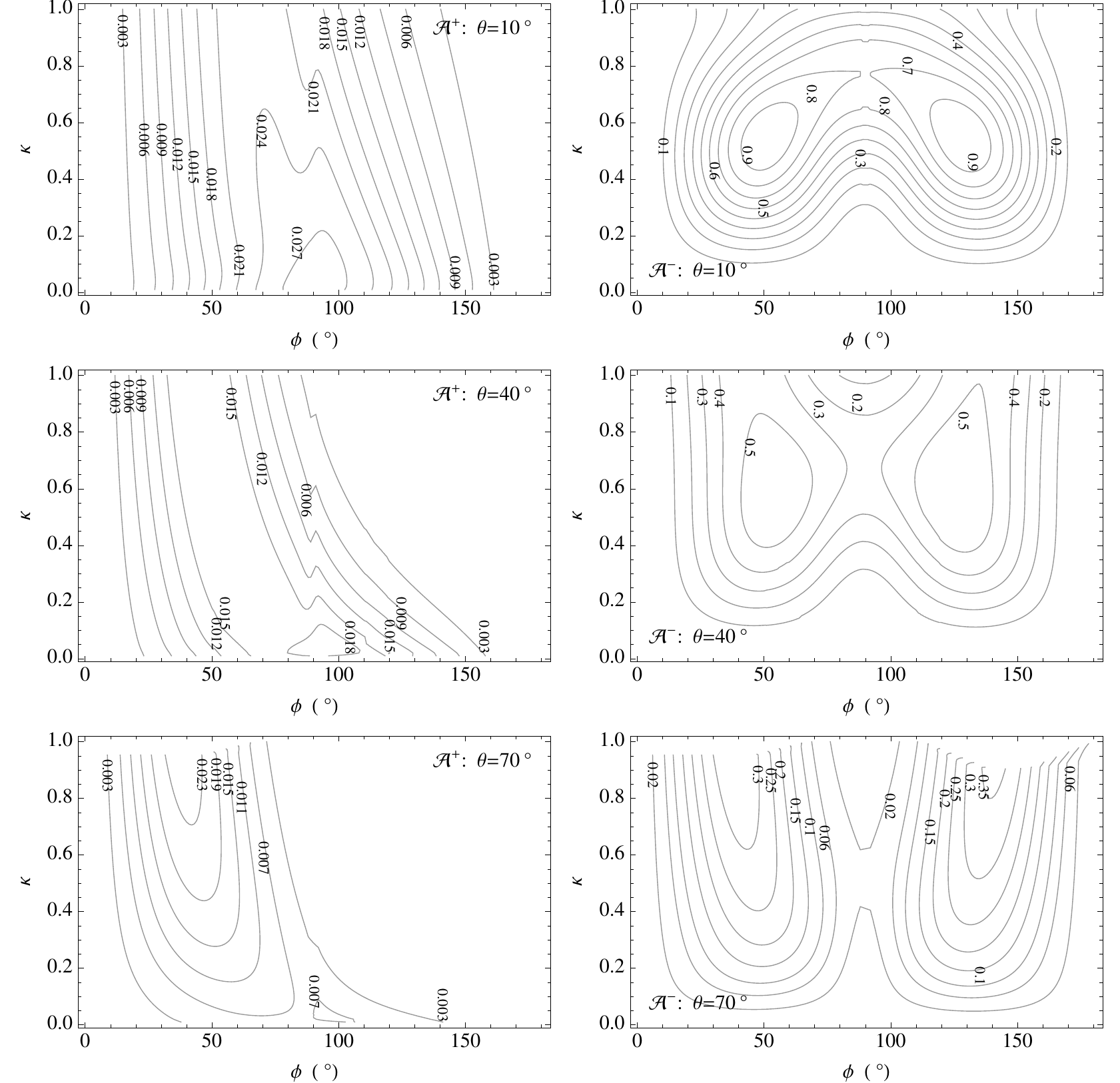}
\caption{Left column: The forward Alfv\'en conversion coefficient $\mathscr{A}^+$ as a function of $\phi$ and $\kappa$ for $\theta=10^\circ$, $40^\circ$, and $70^\circ$ (top to bottom) as labelled with $h_{\rm cor}=100$ throughout. Right column: The reverse Alfv\'en conversion coefficient $\mathscr{A}^-$ for the same cases. In all cases, the transition region is at $x_T = 8$. 
The respective maxima in $\mathscr{A}^+$ are: 0.028 for $\theta=10^\circ$;  
 0.019 for $\theta=40^\circ$;
 0.023 for $\theta=70^\circ$.  The bottom two panels, displaying the results for $\theta=70^\circ$, are truncated near the top for numerical reasons.  This is of little consequence as physically we are mainly interested in $\kappa\lesssim0.2$.
}
\label{fig:conts10to70}
\end{center}
\end{figure*}

\begin{figure*}[tbhp]
\begin{center}
\includegraphics[width=\hsize]{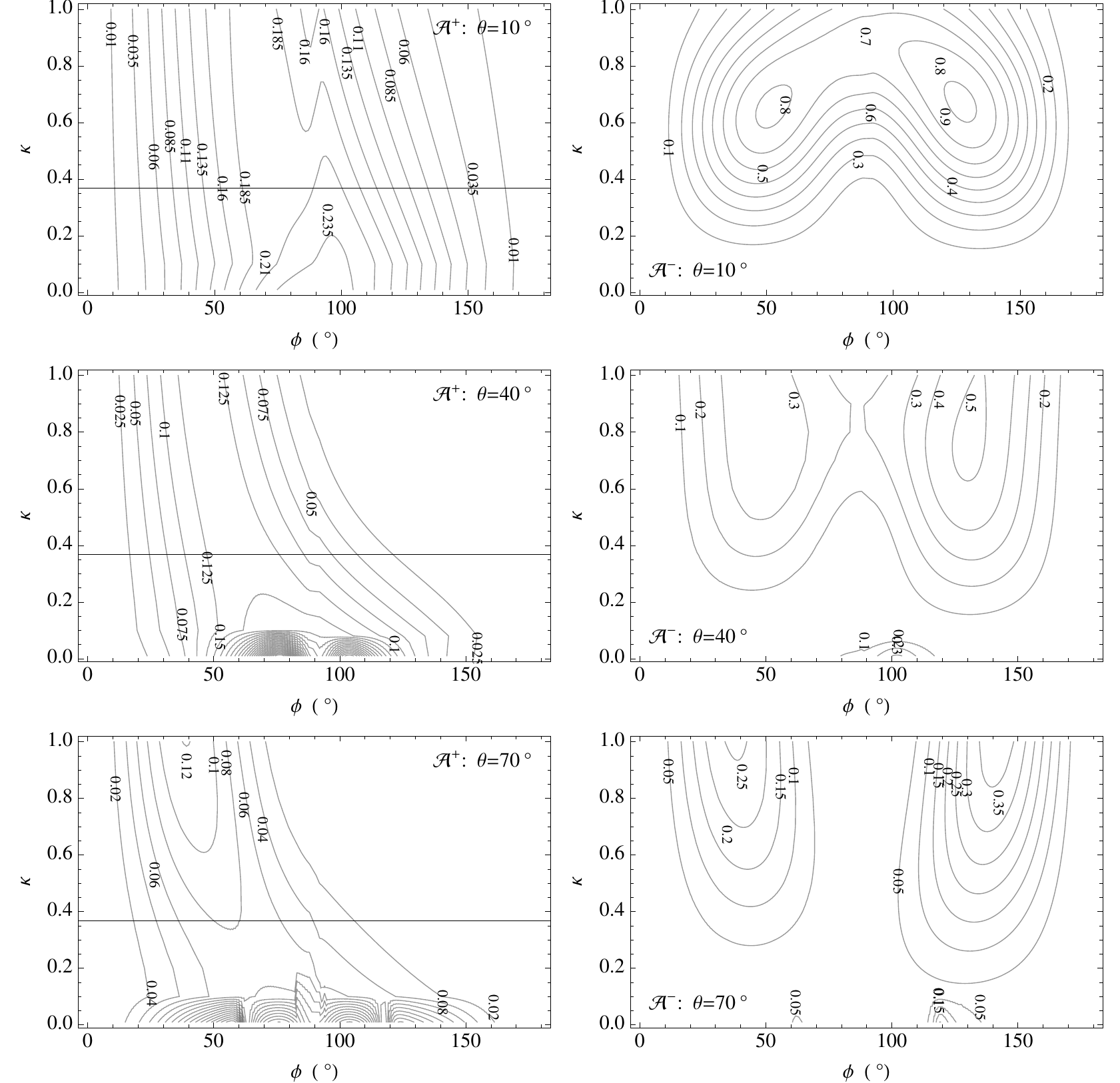}
\caption{Same as Fig.~\ref{fig:conts10to70}, but with $x_T=2$. The horizontal line $\kappa=e^{-x_T/2}=0.368$ in the left panels corresponds to the boundary between the parameter regions where the fast wave reaches the TR before reflecting (below the line) and where it does not (above). Substantially larger values of $\mathscr{A}^+$, exceeding 0.3, are found below the line, for $\theta=40^\circ$.
}
\label{fig:conts10to70xT2}
\end{center}
\end{figure*}

We now fix both $x_T=8$ and $h_{\rm cor}=100$, and explore variation with $\kappa$, $\theta$, and $\phi$. Figure \ref{fig:conts10to70} shows graphically how $\mathscr{A}^+$ and $\mathscr{A}^-$  vary with transverse wavenumber $\kappa$ and wave polarization $\phi$ for magnetic field inclination $\theta = 10^\circ$, $40^\circ$ and $70^\circ$. Across the board, $\mathscr{A}^+$ does not exceed about 2-3\%. The backward Alfv\'en flux $\mathscr{A}^-$ is approximately symmetrical about $\phi\sim90^\circ$.  This is expected in light of the near-symmetry between $\mathscr{A}^+$ and $\mathscr{A}^-$ about $\phi=90^\circ$ found in Paper I for the single layer model. We surmise that converted forward-propagating Alfv\'en waves are near-totally reflected off the transition region when $\phi\lesssim90^\circ$ and when $\phi\gtrsim90^\circ$ the fast wave is partially converted to a backward-propagating  Alfv\'en wave only after fast wave reflection at $x_R$. Maximal backward Alfv\'en flux seems to occur around $50^\circ$ and $140^\circ$ in each case, but typically at unrealistically high $\kappa$ values ($\kappa\gtrsim0.5$). 

Reducing $x_T$ to 2 (Fig.~\ref{fig:conts10to70xT2}) places the TR before (if $\kappa<e^{-x_T/2}=0.368$) or low in the tail of the fast wave reflection point, with the effect of substantially increasing $\mathscr{A}^+$. This suggests that the fast wave evanescent tail, which easily penetrates the TR, continues to produce Alfv\'en waves in the low corona.

\begin{figure*}[tb]
\begin{center}
\includegraphics[width=\hsize]{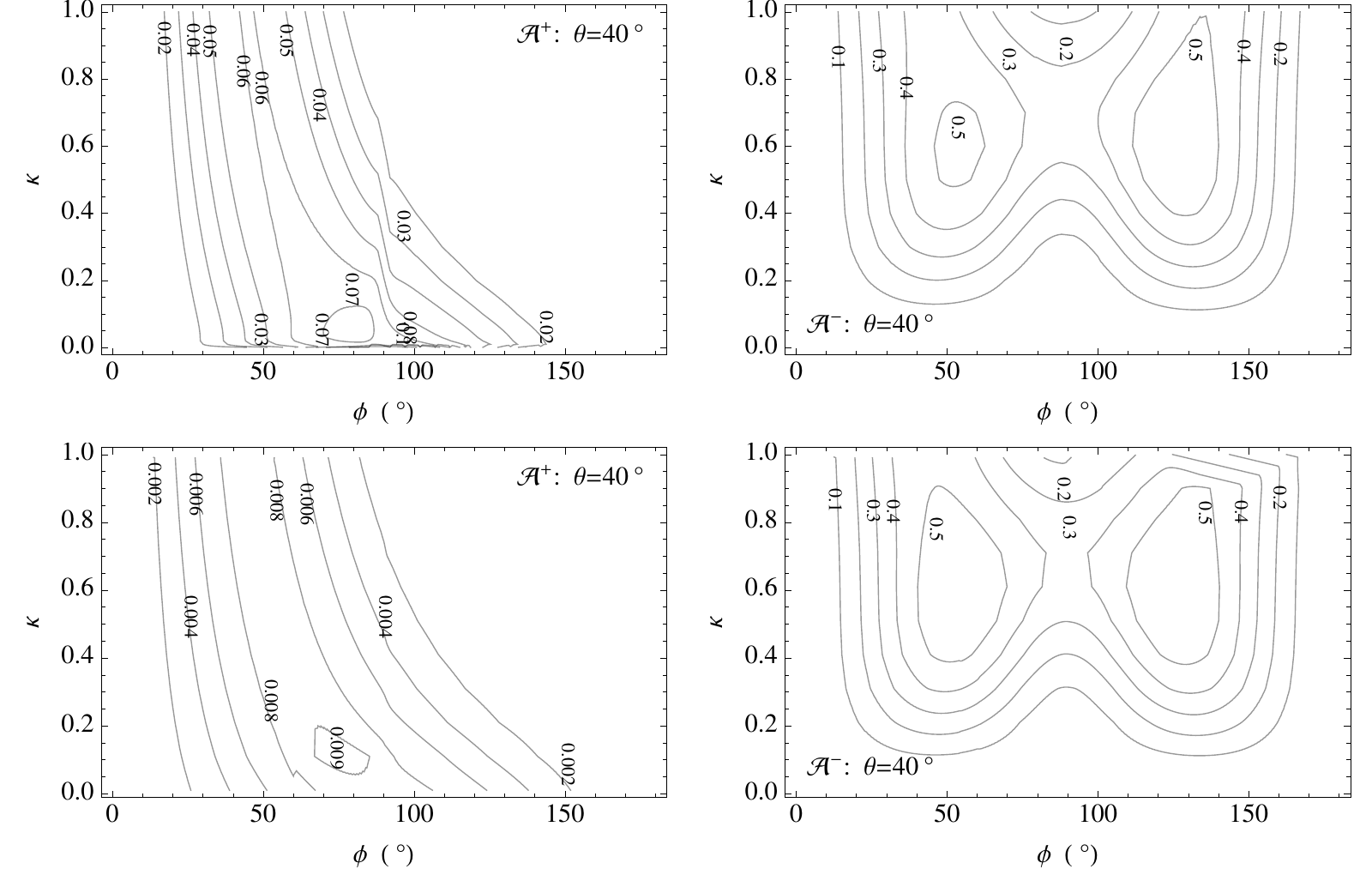}
\caption{Left column: The forward Alfv\'en conversion coefficient $\mathscr{A}^+$ as a function of $\phi$ and $\kappa$ for fixed $\theta=40^\circ$ with $x_T=5$ and $x_T=10$ (top to bottom) and $h_{\rm cor}=100$ throughout. Right column: The reverse Alfv\'en conversion coefficient $\mathscr{A}^-$ for the same cases. }
\label{fig:conts40xT}
\end{center}
\end{figure*}

Finally, Figure \ref{fig:conts40xT} demonstrates how $\mathscr{A}^+$ and $\mathscr{A}^-$  vary with  $\kappa$ and  $\phi$ for fixed $\theta = 40^\circ$ with two different $x_T$ values.  While $\mathscr{A}^+$ is reduced significantly by increasing $x_T$, there is little change to $\mathscr{A}^-$.

%%%%%%
\subsection{Phase Shifts}\label{phase}
With a transition region in place, we have seen that by far the bulk of the injected fast wave flux
is reflected, presumably to once again undergo magnetic-to-acoustic mode conversion at the Alfv\'en-acoustic equipartition level and then re-enter the Sun's helioseismic wave field. Helioseismic techniques such as Time-Distance helioseismology \citep{DuvJefHar93aa} and Phase Sensitive Holography \citep{BraLin00ab} use phase to infer travel times and so are potentially compromised by any anomalous phase shifts suffered by the fast wave in the chromosphere. The cold plasma model affords a particularly convenient testbed for this effect. In such atmospheres the fast wave phase and group speeds are just $a$, independent of the relative orientation of the wavevector and the magnetic field. So, apart from the Alfv\'en conversion mechanism, reflected fast wave phase should be independent of $\theta$ and $\phi$. (Specifically, this is apparent from Equation (3) of Paper I with $k_y$ set to zero. Then the fast and Alfv\'en waves decouple and the fast wave is governed simply by $(\nabla^2+\omega^2/a^2)\xi_\perp=0$, totally independent of $\theta$.) We confirm numerically that indeed the phase of the returning fast wave relative to its injection phase is independent of $\theta$ when $k_y=0$. In this section we investigate phase in both the single-layer model and the model with transition region by comparing the phase of the returning fast wave with that for the $\phi=0$ case where the fast and Alfv\'en waves are entirely decoupled.

\begin{figure*}[tbhp]
\begin{center}
\includegraphics[width=\hsize]{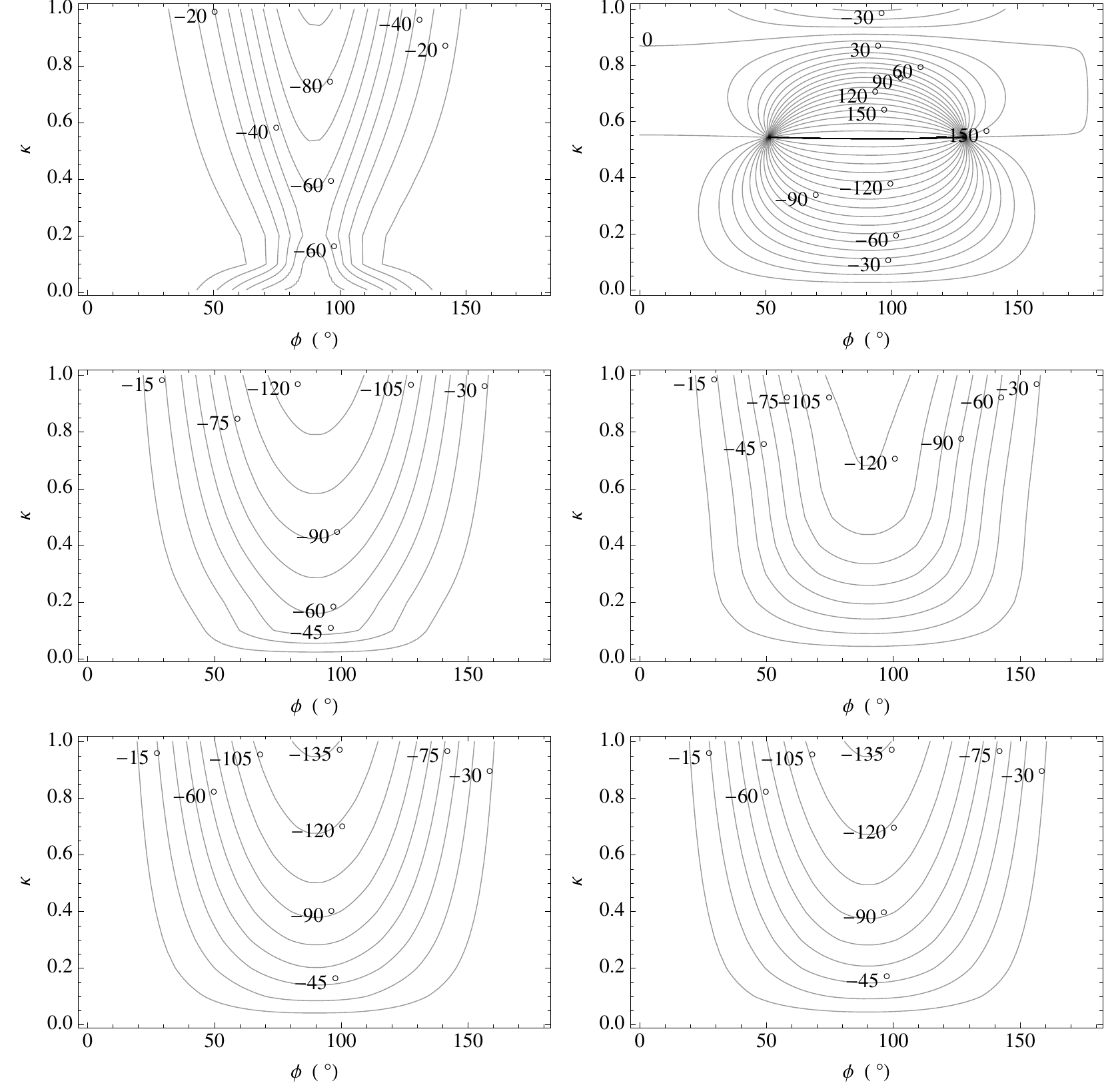}
\caption{The phase difference $\Delta\varphi$ as a function of $\phi$ and $\kappa$ for $\theta=10^\circ$, $40^\circ$, and $70^\circ$ (top to bottom).  Left column: no transition region; Right column: transition region is at $x_T = 8$ with $h_{\rm cor}=100$. The `cut' in the top right panel results from the phases being restricted to $-180^\circ<\Delta\varphi\le180^\circ$.}
\label{fig:contsphase10to70}
\end{center}
\end{figure*}

The fast wave is most conveniently characterised by the dilatation, $\chi=\Div\bxi$ and thus the forward and backward fast wave dilatations $\chi^+$ and $\chi^-$ carry information about their respective phases.  The phase change at any point $x$ is $\delta \varphi = -\arg(\chi^+ \chi^-)$. However, we require the difference between the magnetic and non-magnetic cases $\Delta \varphi=\delta\varphi-\delta\varphi_0$ \citep{Cal09ab}, where $\delta\varphi_0$ refers to the $\phi=0^\circ$ case, which has no Alfv\'en interaction. This quantity is independent of depth provided it is calculated well below the fast wave reflection point at $x=x_R$.  We therefore choose the depth where the WKB solution is applied as the point to evaluate the phase difference.\footnote{For completeness, we mention the result from Paper I \citep[see also][]{FerPlu58aa} that the decoupled ($k_y=0$) fast wave has exact solution $\xi_\perp=a\,J_{2\kappa}(2e^{-x/h})$, with arbitrary complex constant $a$. This splits into forward and backward travelling wave components: $\xi_\perp^\pm=\half a\,H_{2\kappa}^{(2,1)}(2e^{-x/h})$. Hence $\chi^\pm=\partial_\perp\xi_\perp=-\half a[i\,\kappa\,e^{i\,\theta}H^{(2,1)}_{2\kappa}(2e^{-x/h})+e^{-x/2h}H^{(2,1)}_{2\kappa-1}(2e^{-x/h})]$. Consequently,
$\delta\varphi_0\to-2\arg a$ as $x\to-\infty$. This cancels in $\Delta\varphi$ with an equivalent term in $\delta\varphi$ to leave only an effect related to the Alfv\'enic mode conversion.}  Figure  \ref{fig:contsphase10to70} shows the phase difference $-180^\circ < \Delta\varphi \leq 180^\circ$  as a function of $\phi$ and $\kappa$ for fixed $h_{\rm cor}=100$, $x_T=8$ and $\theta=10^\circ$, $40^\circ$ and $70^\circ$, displaying symmetry about $\phi=90^\circ$. The maximum phase difference occurs for higher $\kappa$ as $\theta$ increases.  The phase difference is generally maximal around $\phi=90^\circ$.

\begin{figure*}[tbhp]
\begin{center}
\includegraphics[width=.6\hsize]{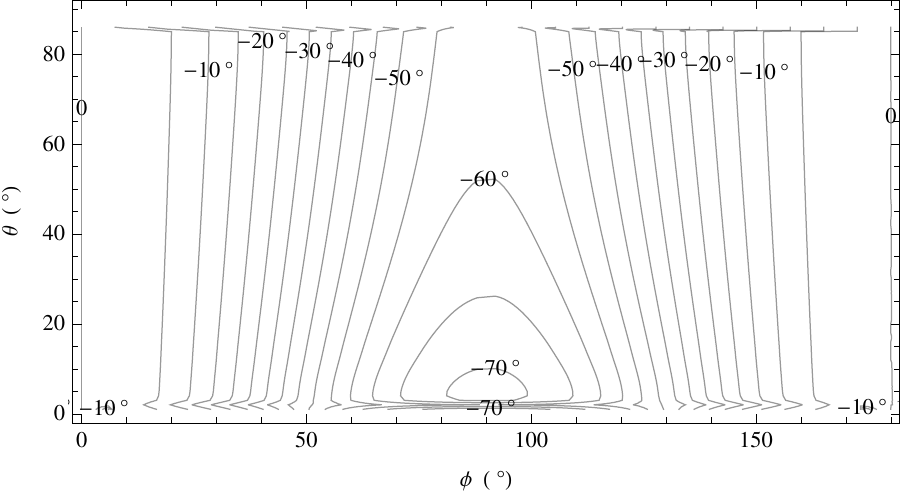}
\caption{Phase $\Delta\varphi$ against $\phi$ and $\theta$ for fixed $\kappa=0.2$ for the case with transition region at $x_T=8$ with $h_{\rm cor}=100$. By definition, $\Delta\varphi=0$ for all $\theta$ at $\phi=0^\circ$ and $180^\circ$.}
\label{fig:contphasekappa}
\end{center}
\end{figure*}

Finally, Fig.~\ref{fig:contphasekappa} demonstrates that for the most part, as expected, phase shift depends only weakly on field inclination $\theta$. However, this is not the case for $\phi\approx90^\circ$.

Overall then, it is clear that the phase of the returning fast wave has been very substantially altered by mode conversion. Without this interaction with the Alfv\'en wave, $\Delta\varphi$ should be identically zero for all $\phi$ and $\theta$. It most certainly is not zero! Any phase sensitive seismology performed using the returning fast wave is therefore severely compromised by this highly directional effect. 

%%%%%%%%%%%%%%%%%%%
\section{Discussion and Conclusions} 

We are motivated by doubt about where the Alfv\'en waves  that are observed in the chromosphere, corona and solar wind are generated.  There is reason to believe \citep{Par91aa,VraPoePan08aa} that the creation of these waves in the photosphere is problematic.  Irrespective of that, it appears from Paper I and  \citet{KhoCal12aa} that it is easy to generate substantial Alfv\'enic flux in the chromosphere.  The total outgoing Alfv\'en conversion coefficient $\mathscr{A}^+$ is sensitive to the distance between the fast wave reflection point and the TR. If $x_T$ is small, substantial Alfv\'en flux can get through, up to about a third, for $h_{\rm cor}$ of 100.  Increasing the distance results in stronger reflection of these waves.

As we see in Figure  \ref{fig:contsphase10to70} increasing the magnetic field inclination reduces  the impact of the TR on phase shifts.  This may be due to increasing Alfv\'en path length along the oblique field lines. {For the most part, the phase shifts calculated here and displayed in Figures \ref{fig:contsphase10to70} and \ref{fig:contphasekappa} are negative, corresponding to a \emph{delay} in the phase of the reflected fast wave. This could be misinterpreted as an increase in travel time. We interpret it instead as an artefact of mode conversion.}

How much energy is carried by the observed coronal Alfv\'en waves?  \citet{TomMcIKei07aa} calculate an insignificant Alfv\'en flux of $0.01$ W m$^{-2}$ though they admit this may be gross underestimate due to insufficient spatial resolution.  Recently this figure has been revised dramatically upward by \citet{McIde-Car11aa} using the He \textsc{ii} 304{\AA} and Fe \textsc{ix} 171{\AA} channels of the Atmospheric Imaging Assembly (AIA) aboard the Solar Dynamics Observatory (SDO) to achieve arcsecond resolution. They estimate energy fluxes of $\sim 100$ W\,m$^{-2}$ in active region loops based on typical amplitudes of around $20$ km\,s$^{-1}$.  

In the most ideal case (large attack angle), almost all flux through the $a=c$ level converts to the fast wave. Of this, at most $30\%$ is carried through the TR by the Alfv\'en wave, but only if the fast wave reaches or nearly reaches the TR before reflecting.  Based on the photospheric p-mode power distribution of \citet{Tho85aa} obtained using the Fe \textsc{i} 6303 line, which is formed about 290 km above optical depth unity, and the VAL C empirical model of \citet{VerAvrLoe81aa}, we crudely estimate $\langle v^2\rangle\sim 4\times10^3$ m$^{2}$\,s$^{-2}$ and a flux of order $800$ W\,m$^{-2}$ associated with p-modes in the $3-5$ mHz band.  This appears to be sufficient to supply the coronal Alfv\'en flux estimated by \citet{McIde-Car11aa},
though the uncertainties in each calculation are large. The link is plausible at least. {Of course, one should not rule out the possibility of multiple sources of coronal Alfv\'en waves, including direct photospheric excitation and chromospheric fast wave conversion.} 

%We have previously noted the effect ambipolar diffusion due to low ionization fraction and high ion-neutral collision frequency has on suppressing the generation of Alfv\'en waves in the photosphere, and partially in response to this postulated that instead fast MHD waves might generate Alfv\'en waves high in the chromosphere where partial ionization is not important. However, this raises the question of how well the fast waves themselves might penetrate the ambipolar layer. It is known from studies related to molecular clouds that, dependent on wavelength, the dissipational effects can be profound \citep{vanFalHar08aa,MouCioMor11aa}. In the solar context though \citep{KhoArbRuc04aa,KhoRucOli06aa,VraPoePan08aa,ZaqKhoRuc11aa}, the ambipolar layer is quite thin relative to the wavelength of the relevant oscillations, and so we might expect the fast wave to survive the transit. Specifically, judging by Table 2 of \citet{VraPoePan08aa}, the ratio of imaginary to real parts of the frequencies derived from a dispersion relation is quite tiny for wavelengths characteristic of seismic waves (several Mm or more). 

%Where the Alfv\'en-acoustic equipartition layer lies higher in the atmosphere than the ambipolar layer, as in network say, the fast waves themselves do not need to traverse the high dissipation region, so there is no issue.

%\acknowledgements

%%%%%%%%%%%%%
%\clearpage
\appendix
\section{On the Unidirectionality of Pure Alfv\'en Waves in an Exponential Atmosphere}  \label{uni}

Decoupled Alfv\'en waves in the case $k_y=0$ admit exact solutions (\ref{hankel}) in terms of
Hankel functions $H_0^{(1)}$ and $H_0^{(2)}$. This is because the Alfv\'en wave equation $\partial^2\xi_y/\partial t^2=a^2\,\partial^2\xi_y/\partial s^2$ (where $s$ is distance along a field line) transforms into the axisymmetric wave equation on a \emph{uniform} membrane 
\begin{equation}
\pderivd{\xi_y}{t} = \frac{\omega^2}{r} \pderiv{}{r}\left( r \pderiv{\xi_y}{r} \right)
\end{equation}
under the change of variables $r = 2 \omega h/a_x$, where $a_x=a\cos\theta$. The Alfv\'en wave 
in the exponential atmosphere therefore is isomorphic to the problem of axisymmetric 2D waves, with $x=+\infty$ mapping to $r=0$. As such, there is no reflection, despite expectations that the exponential atmosphere might be at least partially reflective to Alfv\'en waves. As shown by \citet[Section 7.4]{Whi74aa}, the 2D axisymmetric wave equation with an axial harmonic source results in the $H_0^{(1)}(r)$ solution \citep[see also][p.~194]{CouHil62aa}. We are concerned with the time reverse, or equivalently complex conjugate, where $x=+\infty$ is a sink, selecting $H_0^{(2)}(r)$ instead. See Equation (\ref{Poynting}) for the wave energy fluxes associated with the two Hankel functions. As a further demonstration that $H_0^{(2)}(r)$ corresponds entirely to wave propagation in the negative $r$-direction (positive $x$), we manipulate Equation (9.1.25) of \cite{AbrSte65aa} to show that
\begin{equation}
\begin{split}
H_0^{(2)}(r) = \frac{i}{\pi} \int_{-\infty}^{\infty-i\pi} e^{r\sinh\tau}d\tau &=
 \frac{2}{\pi} \int_0^1\frac{e^{-i\,k\,r}}{\sqrt{1-k^2}} \,dk +\frac{2i}{\pi}\int_0^\infty e^{-r\sinh\tau}d\tau \\[8pt]
 &= \alpha(r) + i\, \beta(r)
 \, ,  
 \end{split}
 \label{Hankel2}
\end{equation}
clearly demonstrating that there are only Fourier components associated with negative $r$ propagation. The first term $\alpha(r)=J_0(r)-i\,\mathbf{H}_0(r)$ is a complex Fourier integral containing only waves propagating towards $r=0$.
The second term $\beta(r)=\mathbf{H}_0(r)-Y_0(r)$, is of course real, positive, and monotonic decreasing in $r$, and so is not wavelike at all. In these expressions, $J_0$ and $Y_0$ are the Bessel functions of the first and second kind, and $\mathbf{H}_0$ is the Struve function. Transmission of Alfv\'en waves in an infinite exponential atmosphere is therefore total, $\mathscr{T}=1$, though discontinuities such as we use to model the TR produce substantial reflection (see Equation (\ref{T})).

\begin{figure}[thb]
\begin{center}
\includegraphics[width=0.6\hsize]{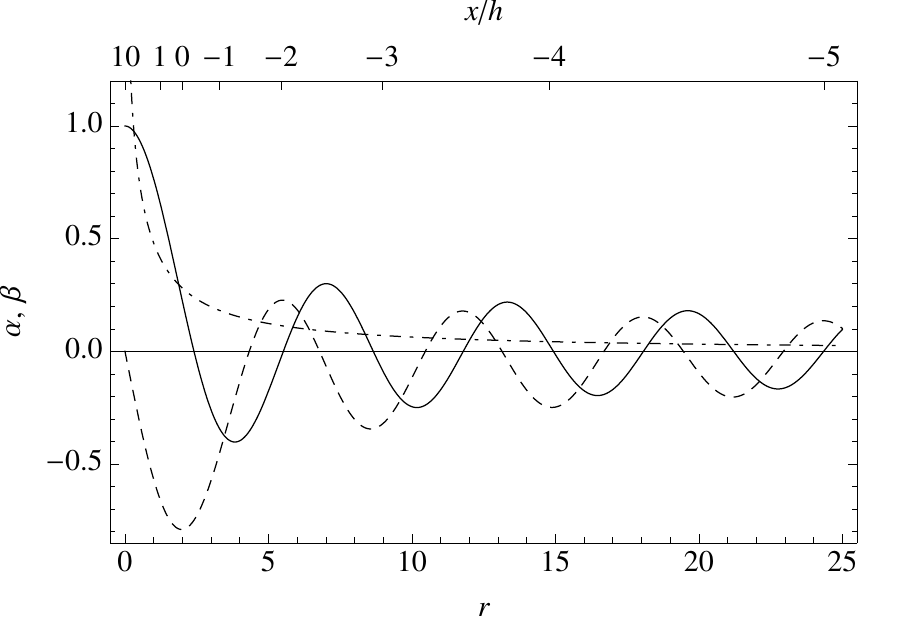}
\caption{The wavelike part $\alpha$ of $H_0^{(2)}(r)$ (real part as full curve, imaginary part as dashed curve) and the `reverberation' part $\beta$ (chained curve), all plotted as functions of $r=2 \exp[-x/2h]$.}
\label{fig:alphabeta}
\end{center}
\end{figure}

\cite{HolIse07aa} examined the behaviour of an Alfv\'en pulse in a stratified atmosphere and inferred partial reflection in a wake behind the propagating packet. However, we interpret this a little differently. It is well known that waves in spaces of even spatial dimension do not obey Huygens' principle, meaning that any wavefront trails a wake that ``persists there indefinitely as a {`reverberation'}\,'' in the words of \citet[pp.~208--210]{CouHil62aa}. Hence, since the Alfv\'en problem is isomorphic to the 2D wave equation, that wake structure is to be expected. We see it as represented by $\beta$. It need not be interpreted as a reflection, though of course, one could express $\beta$ as a Fourier integral, 
$
\beta=2\,\pi^{-1}\left(
\int_0^{\pi/2} \sin(r\cos\theta)\,d\theta +
\int_0^\infty \cos(r \cosh\tau)\,d\tau
\right)
$
representing the reverberation as a sum of standing waves, consisting of equal and opposite forward and backward parts. This is consistent with the view taken by \cite{HolIse07aa}. The important point though is that the reflection, if it is interpreted as such, is spatially confined to the wake and does not include a finite reflection of energy back to $x=-\infty$. An alternate representation of the wake term $\beta$ though is the real integral displayed in Equation (\ref{Hankel2}), or equivalently the Laplace integral 
$
\beta=2\,\pi^{-1} \int_0^\infty e^{-r \tau}(1+\tau^2)^{-1/2}\,d\tau
$.
This is an equally valid representation that suggests an in-place oscillation, characteristic of solutions of the wave equation in even-dimensional spaces. 

If we turn on a harmonic driver at a particular time, we expect the outermost wavefront to trail a wake which develops into the $\beta$ term. Once steady oscillations are set up, after the transient has passed, all energy is propagated to infinity. The wavelike and reverberation parts of $H_0^{(2)}(r)$ are plotted in Figure \ref{fig:alphabeta}, showing that the latter are most significant at small $r$ (large $x$). Asymptotically, $\beta\sim x/h\pi$ as $x\to+\infty$, so the reverberation takes the form of a straight `flapping' of magnetic field lines, as seen in the animations accompanying \citet{CalHan11aa}.

%\clearpage

%%%%%%%%%%%%%%%%%%%%%%%%%%%%%%%
%  REFERENCES

\bibliographystyle{apj}        
\bibliography{fred}

\end{document}